\documentclass[11pt,twoside, final]{amsart}
\copyrightinfo{0}{Iranian Mathematical Society}
\usepackage{amsmath,amsthm,amscd,amsfonts,amssymb,enumerate}
\usepackage{graphicx}		
\usepackage{color}
\usepackage[colorlinks]{hyperref}
\newtheorem{theorem}{Theorem}[section]
\newtheorem{proposition}[theorem]{Proposition}

\theoremstyle{definition}

\theoremstyle{remark}

\numberwithin{equation}{section}
\commby{}
\begin{document}

\begin{center}  \tiny{To Appear in}\end{center}\begin{center}\tiny{The Bulletin of the Iranian Mathematical Society (BIMS)}\end{center}$\\\\$
\title[Trivially Related Lax Pairs Of The Sawada-Kotera Equation]{Trivially Related Lax Pairs Of The Sawada-Kotera Equation}
\author[D. Talati]{Daryoush TALATI}
\address{Department of Engineering Physics, Ankara University 06100 Tando\u{g}an$\\$~Ankara}
\email{talati@eng.ankara.edu.tr, daryoush.talati@gmail.com}
\maketitle
\begin{abstract}
We show that a recently introduced Lax pair of the Sawada-Kotera equation is not a new one but is trivially related to the known old Lax pair. Using the so-called trivial compositions of the old Lax pairs with a differentially constrained arbitrary operators, we give some examples of trivial Lax pairs of KdV and Sawada-Kotera equations.
\end{abstract}
\section{\bf Introduction}
The term ‘Lax pair’ refers to linear systems that are related to nonlinear equations through a compatibility condition. The first part of Lax pair is called the scattering problem, that allows the initial-value problem for the integrable equation to be solved exactly. If a nonlinear equation possesses a Lax pair, then the Lax pair may be used to gather information about the behavior of the solutions to the nonlinear equation. Importantly, the existence of a Lax pair is a signature of integrability of the associated nonlinear equation. In his seminal work \cite{lax}, Lax suggested a formalism to integrate a class of nonlinear evolution equations. He introduced a pair of linear operators $L$ and $M$ such that
\begin{equation} \begin{array}{ll}
L\phi=\lambda \phi,\\
\phi_t=M\phi,\label{123}
\end{array} \end{equation}
where L and M are linear differential operators, $\lambda$ is an eigenvalue of $L$, and $\phi$ is an eigenfunction of $L$. Assuming $\lambda_t = 0$, differentiating $L\phi$ with respect to t gives
\begin{equation}
L_t\phi+L\phi_t=\lambda\phi_t.\label{isp}
\end{equation}
Substituting in from \eqref{123} gives that
\begin{equation}
L_t\phi=ML\phi-LM\phi.
\end{equation}
By an explicit computation we have
\begin{equation}
(L_t+[L,M])\phi=0,
\end{equation}
where $[M,L]= ML - LM$ is the operator commutator. Hence
\begin{equation}
L_t=[M,L],
\end{equation}
is called Lax equation and contains commutative nonlinear evolution equation for suitable $L$
and $M$. Consider the Lax formalism for the KdV equation
\begin{equation}\begin{array}{ll}
L= D_{x}^2+u+ u_{x}D_{x}^{-1},\\
M=-4D_{x}^3-6 uD_{x}-9u_{x}-3 u_{xx}D_{x}^{-1}.\label{kdv2}
\end{array}\end{equation}
These operators satisfies the Lax equation
\begin{equation}\begin{array}{ll}
~L_t=u_t+ u_{tx}D_{x}^{-1},\\~[M,L]=-(u_{xxxx}+6uu_{xx}+6u_{x}^2)D_{x}^{-1}- u_{xxx}-6uu_{x},
\end{array}\end{equation}
if $u$ is solution to the KdV equation
\begin{equation}\begin{array}{l}
u_t+ u_{xxx}+6uu_{x}=0.
\end{array}\end{equation}
A second Lax pair for the KdV equation is
\begin{equation}\begin{array}{ll}
\grave{L}&= D_{x}^2+u, \\
\grave{M}&=-4D_{x}^3-6uD_{x}-3u_{x}. \label{kdv1}
\end{array}\end{equation}
Recently, a new Lax pair was obtained by Hickman et al. \cite{here}. It is shown that the Sawada-Kotera equation \cite{sawada}
\begin{equation}
u_t +u_{xxxxx}+5uu_{xxx}+5u_{x}u_{xx}+5u^2u_{x} =0,
\end{equation}
possesses two different Lax pairs
\begin{equation}
\begin{array}{ll}
L_1&=D_{x}^3 + uD_{x},\\
M_1&=9D_{x}^5 +15uD_{x}^3 + 15u_{x}D_{x}^2+(5u^2+10u_{xx})D_x, \label{lax1}
\end{array}\end{equation}
and also
\begin{equation}\begin{array}{ll}
L_2&=D_{x}^3 + uD_{x}+u_{x},\\
M_2&=9D_{x}^5 +15uD_{x}^3 + 30u_{x}D_{x}^2+(5u^2+25u_{xx})D_{x}\\&~~~~~~ + 10uu_{x} + 10u_{xxx}, \label{lax2}
\end{array}\end{equation}
such that related compatibility conditions are
\begin{equation}\begin{array}{ll}
L_{1_t}+[L_1,M_1]=0,\\L_{2_t}+[L_2,M_2]=0.
\end{array}\end{equation}
The first Lax pair \eqref{lax1} is well known \cite{lax11}. The second Lax pair is new. It appeared in \cite{here} for the first time. Later in \cite{sako} using the method of gauge-invariant description of zero-curvature representations (ZCRs) and the method of cyclic bases of ZCRs, it is shown that the new Lax pair \eqref{lax2} is equivalent to the well-known old Lax pair of this equation only if the Lax pairs are considered in the form of ZCRs.
\section{\bf Trivially related Lax pairs}
This work discusses the generation of an infinite number of Lax pairs for a nonlinear nonlinear equation using one known Lax pair. A Lax pair $(\mathcal{L}_2,\mathcal{M}_2)$ which is obtainable from other compatible Lax pairs $(\mathcal{L}_1,\mathcal{M}_1)$ as $ \mathcal{L}_2=O\mathcal{L}_1O^{-1}$ and $\mathcal{M}_2=O\mathcal{M}_1O^{-1}$ is trivially related since the Lax pair $(\mathcal{L}_2,\mathcal{M}_2)$ gives a subset of the structure that the pair $(\mathcal{L}_1,\mathcal{M}_1)$ gives. Taking the product on the left with $t$-independent operator $O$ and on the right with operator $O^{-1}$, the compatibility condition \eqref{isp} can be expressed directly in terms of the operators $\mathcal{L}_1,\mathcal{M}_1,O$ and $O^{-1}$. Indeed, from these we have
\begin{equation}
O\mathcal{L}_{1_t}O^{-1} =O(\mathcal{M}_{1}\mathcal{L}_{1} -\mathcal{L}_{1}\mathcal{M}_{1})O^{-1}.
\end{equation}
By an explicit computation we get
\begin{equation}
O\mathcal{L}_{1_t}O^{-1} =O \mathcal{M}_{1}O^{-1}O\mathcal{L}_{1}O^{-1} - O\mathcal{L}_{1}O^{-1}O\mathcal{M}_{1} O^{-1}.
\end{equation}
Hence, using $[\frac{d}{dt},O]=0$  we arrive at the following formula:
\begin{equation}
(\underbrace{O\mathcal{L}_1O^{-1}}_{\mathcal{L}_2})_t =(\underbrace{O\mathcal{M}_1O^{-1}}_{\mathcal{M}_2})(\underbrace{O\mathcal{L}_1O^{-1}}_{\mathcal{L}_2}) - (\underbrace{O\mathcal{L}_1O^{-1}}_{\mathcal{L}_2})(\underbrace{O\mathcal{M}_1 O^{-1}}_{\mathcal{M}_2}).
\end{equation}
So $(\mathcal{L}_2,\mathcal{M}_2)$ is a Lax pair, and it satisfies the compatibility condition
\begin{equation}\begin{array}{ll}
\mathcal{L}_{2_t}=[\mathcal{M}_2,\mathcal{L}_2].\label{theo}
\end{array}\end{equation}
However, such models have been rediscovered again and again in the literature. So we consider it meaningful to state these pairs as a proposition.
\begin{proposition}
If a PDE $F(x,u,u_t,u_x, u_{xx}...,u_{nx})=0$ admits a Lax pair $(\mathcal{L}_1,\mathcal{M}_1 )$ then it admits the infinite sequence of trivial Lax pairs  \cite{lax} 
\begin{equation}\begin{array}{ll}
\mathcal{L}_2&=O\mathcal{L}_1O^{-1}~,\\
\mathcal{M}_2&=O\mathcal{M}_1O^{-1},
\end{array}\end{equation}
which share the same set of the structure. $O$ is a $t$-independent operator.
\end{proposition}
Therefore, by straightforward calculation it is easy to show that in \cite{here} the obtained new Lax pair \eqref{lax2} of the Sawada-Kotera equation is trivial compositions of old Lax pair \eqref{lax1} and $O=D_x$. This implies that
\begin{equation}\begin{array}{ll}
L_{2}=D_xL_1D_x^{-1},\\M_{2}=D_xM_1D_x^{-1}.\label{triviall}
\end{array}\end{equation}
It is interesting that according to the arXiv records, just one day after online publication of this work the same observation \eqref{triviall} was made by the author of \cite{sako} too. With $O=D_x^2$, we can generate another trivial Lax pair for Sawada-Kotera equation:
\begin{equation*}
\begin{array}{ll}
L_3&=D_x^2L_1{D_{x}^{-2}}\\& =D_{x}^3+ uD_{x} +2u_{x}+ u_{xx}D_{x}^{-1},\\\\
M_3&=D_x^2M_1{D_x^{-2}} \\&=9D_{x}^5+15 uD_{x}^3+45 u_{x}D_{x}^2+5 (11u_{xx}+u^2)D_{x}\\&+5(7u_{xxx}+4u_{x}u) +10 (u_{xxxx}+u_{xx}u+u_{x}^2)D_{x}^{-1}.
\end{array}\end{equation*}
Another example can be the KdV equation. For the same reasons, the Lax pair \eqref{kdv2} is trivial compositions of old Lax pair \eqref{kdv1} and $O=D_x$ :
\begin{eqnarray}
\begin{array}{ll}
L&= D_x\grave{L}D_x^{-1}\\& =D_x( D_{x}^2+u)D_x^{-1}\\& = D_{x}^2+u+u_{x}D_{x}^{-1},\\\\
M&=D_x\grave{M}D_x^{-1}\\& = -D_x(4D_{x}^3+6uD_{x}+3u_{x}) D_x^{-1}\\& =-(4D_{x}^3+6 uD_{x}+9u_{x}+3 u_{xx}D_{x}^{-1}).
\end{array}
\end{eqnarray}
\section{\bf Discussion}
We have shown that the recently introduced Lax pair of the Sawada-Kotera equation is not a new one.
The new Lax pair can be derived from the old one by taking the product on the old Lax pair left with operator $D_x$ and on the right with operator $D_x^{-1}$. Therefore, we proved that the new Lax pair is trivially related to the known old Lax pair. Such models have been rediscovered again and again in the literature. So we consider it meaningful to state these pairs as a proposition and also some examples of trivial Lax pairs of KdV and Sawada-Kotera equations are given.
\section{\bf Acknowledgements}
The author would like to thank the referees for careful reading, many useful comments and a number of helpful suggestions for improvement in the article.

\end{document}